\documentclass[iop,apj]{emulateapj}
\usepackage{amsmath,amssymb}

\usepackage{xcolor}

\newcommand{\bbv}[1]{\boldsymbol{#1}}
\newcommand{\kfrac}[2]{\frac{\displaystyle{#1}}{\displaystyle{#2}}}

\slugcomment{}

\usepackage[bookmarksnumbered,
colorlinks,
linkcolor=blue,
citecolor=black,
filecolor=black,
urlcolor=blue,
breaklinks=true,
]{hyperref}
\shortauthors{Knobel et al.}
\begin{document}

\slugcomment{The Astrophysical Journal, 755:48 (12pp), 2012 August 10}

\title{A Group-Galaxy cross-correlation function analysis in zCOSMOS\footnotemark[1]}

\author{C.~Knobel\footnotemark[2],
S.~J.~Lilly\footnotemark[2],
%
%enablers
%--------
C.~M.~Carollo\footnotemark[2],
T.~Contini\footnotemark[3,4],
J.-P.~Kneib\footnotemark[5],
O.~Le Fevre\footnotemark[5],
V.~Mainieri\footnotemark[6],
A.~Renzini\footnotemark[7],
M.~Scodeggio\footnotemark[8],
G.~Zamorani\footnotemark[9],
%
%Core team A
%-----------
S.~Bardelli\footnotemark[9],
M.~Bolzonella\footnotemark[9],
A.~Bongiorno\footnotemark[10],
K.~Caputi\footnotemark[2,21],
O.~Cucciati\footnotemark[11],
S.~de la Torre\footnotemark[12],
L.~de Ravel\footnotemark[12],
P.~Franzetti\footnotemark[8],
B.~Garilli\footnotemark[8],
A.~Iovino\footnotemark[11],
P.~Kampczyk\footnotemark[2],
K.~Kova\v{c}\footnotemark[2,18],
F.~Lamareille\footnotemark[3,4],
J.-F.~Le Borgne\footnotemark[3,4],
V.~Le Brun\footnotemark[5],
C.~Maier\footnotemark[2,20],
M.~Mignoli\footnotemark[9],
R.~Pello\footnotemark[3,4],
Y.~Peng\footnotemark[2],
E.~Perez Montero\footnotemark[3,4,13],
V.~Presotto\footnotemark[11],
J.~Silverman\footnotemark[14],
M.~Tanaka\footnotemark[14],
L.~Tasca\footnotemark[5],
L.~Tresse\footnotemark[5],
D.~Vergani\footnotemark[9,22],
E.~Zucca\footnotemark[9],
%
%Core team B
%-----------
%
L.~Barnes\footnotemark[2],
R.~Bordoloi\footnotemark[2],
A.~Cappi\footnotemark[9],
A.~Cimatti\footnotemark[15],
G.~Coppa\footnotemark[10],
A.~M.~Koekemoer\footnotemark[16],
C.~L\'opez-Sanjuan\footnotemark[5],
H.~J.~McCracken\footnotemark[17],
M.~Moresco\footnotemark[15],
P.~Nair\footnotemark[9],
L.~Pozzetti\footnotemark[9],
and N.~Welikala\footnotemark[19]
}

\footnotetext[1]{European Southern Observatory (ESO), Large Program 175.A-0839}

\affil{$^2$Institute for Astronomy, ETH Zurich, Zurich 8093, Switzerland\\
$^3$Institut de Recherche en Astrophysique et Plan\'etologie, CNRS, 14, avenue Edouard Belin, F-31400 Toulouse, France\\
$^4$IRAP, Universit\'e de Toulouse, UPS-OMP, Toulouse, France\\
$^5$Laboratoire d'Astrophysique de Marseille, CNRS/Aix-Marseille Universit\'e, 38 rue Fr\'ed\'eric Joliot-Curie, 13388, Marseille cedex 13, France\\
$^6$European Southern Observatory, Garching, Germany\\
$^7$INAF-Osservatorio Astronomico di Padova, Vicolo dell'Osservatorio 5, 35122, Padova, Italy\\
$^8$INAF-IASF Milano, Milano, Italy\\
$^9$INAF Osservatorio Astronomico di Bologna, via Ranzani 1, I-40127, Bologna, Italy\\
$^{10}$Max Planck Institut f\"ur Extraterrestrische Physik, Garching, Germany\\
$^{11}$INAF Osservatorio Astronomico di Brera, Milan, Italy\\
$^{12}$Institute for Astronomy, University of Edinburgh, Royal Observatory, Edinburgh, EH93HJ, UK\\
$^{13}$Instituto de Astrofisica de Andalucia, CSIC, Apartado de correos 3004, 18080 Granada, Spain\\
$^{14}$Institute for the Physics and Mathematics of the Universe (IPMU), University of Tokyo, Kashiwanoha 5-1-5, Kashiwa, Chiba 277-8568, Japan\\
$^{15}$Dipartimento di Astronomia, Universit\`a degli Studi di Bologna, Bologna, Italy\\
$^{16}$Space Telescope Science Institute, Baltimore, MD 21218, USA\\
$^{17}$Institut d'Astrophysique de Paris, UMR7095 CNRS, Universit\'e Pierre \& Marie Curie, 75014 Paris, France\\
$^{18}$Max Planck Institut f\"ur Astrophysik, Garching, Germany\\
$^{19}$Insitut d'Astrophysique Spatiale, B\^atiment 121, Universit\'e Paris-Sud XI and CNRS, 91405 Orsay Cedex, France\\
$^{20}$Department of Astronomy, University of Vienna, Tuerkenschanzstrasse 17, 1180 Vienna, Austria\\
$^{21}$Kapteyn Astronomical Institute, University of Groningen, P.O.~Box 800, 9700 AV Groningen, The Netherlands\\
$^{22}$INAF-IASF Bologna, Via P.~Gobetti 101, I-40129 Bologna, Italy
}

\begin{abstract}
We present a group-galaxy cross-correlation analysis using a group catalog produced from the 16,500 spectra from the optical zCOSMOS galaxy survey. Our aim is to perform a consistency test in the redshift range $0.2 \leq z \leq 0.8$ between the clustering strength of the groups and mass estimates that are based on the richness of the groups. We measure the linear bias of the groups by means of a group-galaxy cross-correlation analysis and convert it into mass using the bias-mass relation for a given cosmology, checking the systematic errors using realistic group and galaxy mock catalogs. The measured bias for the zCOSMOS groups increases with group richness as expected by the theory of cosmic structure formation and yields masses that are reasonably consistent with the masses estimated from the richness directly, considering the scatter that is obtained from the 24 mock catalogs. An exception are the richest groups at high redshift (estimated to be more massive than $10^{13.5}\ M_\odot$), for which the measured bias is significantly larger than for any of the 24 mock catalogs (corresponding to a 3$\sigma$ effect), which is attributed to the extremely large structure that is present in the COSMOS field at $z \sim 0.7$. Our results are in general agreement with previous studies that reported unusually strong clustering in the COSMOS field.\\[2mm]
\end{abstract}

%\keywords{Galaxies: groups and clusters: general -- Galaxies: statistics -- Cosmology: observations and large-scale structure of Universe}

\section{Introduction}
\setcounter{footnote}{22}

In the current $\Lambda$CDM paradigm of cosmic structure formation, dark matter (DM) halos are biased tracers of the underlying matter field. That is to say, the autocorrelation function $\xi(r,M)$ of halos of mass $M$ is related to the DM linear correlation function $\xi_{\rm lin}(r)$ as
\begin{equation}
\xi(r,M) =  b^2(M) \:\xi_{\rm lin}(r)
\end{equation}
with $b(M)$ being the linear bias, which is a monotonically increasing function with mass \citep{kaiser1984,bardeen1986,cole1989,mo1996}.
Since both functions $\xi_{\rm lin}(r)$ and $b(M)$ are theoretically well understood for a given cosmology, measuring the correlation function for a sample of halos can be used for several applications. If the masses of the halos are known, it can yield constraints on the underlying cosmology. If the masses are not known, adopting the current favored cosmological model provides information on the typical mass of the halos. If constraints exist for both the mass and the cosmology by means of independent measurements, an analysis of the correlation function allows a consistency test within the current paradigm of structure formation in the universe.

In this paper we perform such a consistency test at redshift $0.2 \leq z \leq 0.8$ by assuming the currently favored $\Lambda$CDM cosmology to be correct and comparing the resulting masses that we obtain from a correlation function analysis with other independent estimates of the masses of the halos.  The sample of DM halos is given by the optical group catalog \citep[][hereafter ``K12'']{knobel2012} which was produced using $\sim \! 16,500$ spectroscopic redshifts (the so-called 20k sample) from the zCOSMOS-bright galaxy survey \citep[][S.J.~Lilly et al. 2012, in preparation]{lilly2007,lilly2009}.  In this context, a group is defined as a set of galaxies occupying the same DM halo and the zCOSMOS group catalog was constructed using a group-finding algorithm that was tuned by comparison to extensive mock galaxy catalogs from the Millennium simulation \citep{kitzbichler2007}.  The operational definition of a DM halo in the Millennium simulation is of a friends-of-friends (FOF) group of DM particles connected with a linking length of $b = 0.2$ times mean interparticle separation, corresponding to structures with a mean overdensity of roughly 200 (see e.g., \citealt{more2011} for a recent discussion). The success of the group-finder in terms of the purity and completeness of the resulting catalog is derived by comparison with these same simulations for which the halo membership is of course known. The high purity of the zCOSMOS catalog guarantees that contaminations from fragmented, over-merged, and spurious groups should be small. 

Unfortunately, it is difficult to directly obtain estimates of the DM masses of the group halos from the optical data, e.g., by means of the virial theorem, since most of the groups have just a few identified members. For this reason, we introduced an estimated mass that was based on the observed richness of the group, corrected for variations in the spatial sampling rate (SSR) of the galaxies, and calibrated against the same simulations.  We referred to this estimated mass as the ``fudge mass''.   The aim of this paper is to examine whether the clustering properties of the groups are consistent with them having these masses in reality.

The most straightforward way to perform the consistency test would be by directly estimating the autocorrelation function of the groups for a given mass range. However, this would require detailed knowledge of the spatial selection function of the groups, which will likely depend on the observed richness.  Clearly a poor group will suffer more from any spatial variation in the spectroscopic sampling than a rich one, which will have been recognized even if several of its members were missed. For this reason, it is preferable to measure the cross-correlation function between groups and galaxies instead, i.e., the group-galaxy cross-correlation function. In this case, only the well understood spatial selection function of the 20k galaxy sample is needed and we can avoid worrying about the more complex spatial selection function of the groups.  

In this paper, we measure the group-galaxy cross-correlation function and the galaxy autocorrelation function to perform a consistency test by estimating the group bias in the linear regime and comparing the results to our richness-calibrated masses. This analysis should not depend on the choice of the ``galaxy'' sample, since the bias of this sample drops out from the analysis. We check this by carrying out the analysis with both magnitude- and volume-limited galaxy samples obtaining consistent results. Furthermore, we will perform the analysis in parallel on simulated mock catalogs in order to test our codes, explore systematics, and to obtain an idea of the impact of cosmic variance.  Not least, we can explore whether the large COSMOS field is consistent with the predictions of the Millennium simulations and/or whether it is representative of other regions of sky. This was previously investigated in a couple of studies  \citep{mccracken2007,meneux2009,kovac2010a,delatorre2010,delatorre2011} with the result that the clustering in the COSMOS field is unusually strong compared with simulations and other surveys.

The group-galaxy cross-correlation function was first measured by \cite{seldner1977} for Abell clusters. In the last decade, it was measured in the local universe for 2dfGRS and SDSS group-galaxy samples \citep{yang2005,berlind2006,mountrichas2007,wang2008} and for DEEP2 at $z \sim 1$ \citep{coil2006}.

This paper is organized as follows. In Section \ref{sec:halo_model}, we briefly review the halo model in the linear regime, which is the theoretical basis of this paper. In Section \ref{sec:selected_sample}, we describe the group and galaxy samples that are used in the analysis. The method of the correlation function estimation is described in Section \ref{sec_estimation_method} and in Section \ref{sec:correlation_functions} we discuss the resulting correlation functions for the actual data and for the mock catalogs. In Section \ref{sec:bias_estimation}, we derive the group masses by means of their bias and compare our results with the mock catalogs. Section \ref{sec:conclusion_correlation_function} concludes the paper and summarizes our findings.

Where needed, the concordance cosmology of the mock catalogs is adopted, i.e., Hubble constant $H_0 = 100 h\; \rm km\:s^{-1}\:Mpc^{-1}$ with $h=0.73$, matter density $\Omega_{\rm m} = 0.25$, cosmological constant $\Omega_\Lambda = 0.75$, spectral index $n_{\rm s} = 1$, and linear fluctuation strength $\sigma_8 = 0.9$.  Although this value of $\sigma_8 = 0.9$ is now thought to be a bit high \citep[e.g.,][]{komatsu2011}, it should be remembered that the point of the paper is to check the consistency with the estimates of halo mass from K12 that were calibrated using these same simulations. Also, the effect of $\sigma_8$ on $\xi(r,M)$ and on $b(M)$ goes in opposite directions (see the discussion in Sect.~\ref{sec:masses}). Throughout this paper, we refer to distances in comoving $h^{-1}$ Mpc and to masses in units of $\log(M/M_\odot)$ explicitly assuming $h = 0.73$. We use the term ``dex'' to express the antilogarithm, i.e., 0.1 dex corresponds to a factor $10^{0.1} \simeq 1.259$.

\section{The Halo model}\label{sec:halo_model}

The principle of using clustering properties of particular objects to infer their DM halo masses is well established in terms of the ``halo model''  (\citealt{peacock2000}, \citealt{seljak2000}, see e.g., \citealt{cooray2002} for a review), and is only briefly reviewed here. The halo model is based on the following assumptions: first, all galaxies reside within DM halos following a certain spherical symmetric density profile, where there is always a galaxy at the center of the halo. Second, the distribution $p(N|M)$ of the number of galaxies in halos of mass $M$, which is called ``halo occupation distribution'' (HOD), depends for a given galaxy sample only on the mass $M$ of the halo. Typical further assumptions are that the galaxy density profile within halos follows that of the DM and that central and satellite galaxies within halos constitute different galaxy populations. For the analysis in this paper, we only need the assumptions that galaxies reside in halos and that the bias of the galaxy population, which comes from the HOD, does not vary greatly within a redshift bin.  

For two samples of galaxies $g$ and $g'$ with comoving number densities $n_{\rm g}(\bf x)$ and $n_{\rm g'}(\bf x)$, respectively, the overdensity for either sample is
\begin{equation}
\delta_{i}(\bbv x) = \frac{n_i(\bbv x)-\bar n_i}{\bar n_i}\:,\quad i \in \left\lbrace g,g'\right\rbrace
\end{equation}
with $\bar n_{i} = \langle n_i(\bf x) \rangle$.  The brackets $\langle \rangle$ denote the cosmic average which is taken to be the volume average by assuming ergodicity or a ``fair sample hypothesis'' for the corresponding density field (see e.g., the discussion in \citealt[][Sect.~4.3]{watts2003}). Since zCOSMOS is too small to constitute a ``fair sample'' especially at small redshift, the corresponding averages are affected by cosmic variance.

The cross-correlation function $\xi_{\rm gg'}(r)$ between these two samples $g$ and $g'$ is defined by
\begin{equation}\label{eq:def_correlation_function}
\xi_{\rm gg'}(r) = \langle \delta_{\rm g}(\bbv x) \delta_{\rm g'}(\bbv x') \rangle = \frac{\langle n_{\rm g}(\bbv x) n_{\rm g'}(\bbv x')\rangle}{\bar n_{\rm g}\bar n_{\rm g'}}-1\:
\end{equation}
with $r = |\bbv x'-\bbv x|$ and is a measure of the excess of $gg'$ pairs at separations $r$ compared to their mean number densities. The reason that $\xi_{\rm gg'}(r)$ depends only on the separation $r = |\bbv x_2-\bbv x_1|$ is a consequence of the basic cosmological assumption of homogeneity and isotropy. 

From the definition of the correlation function and the assumption that all galaxies reside within DM halos, it follows immediately that the cross-correlation function divides into two terms, i.e.,
\begin{equation}
\xi_{\rm gg'}(r) = \xi_{\rm gg'}^{\rm (h1)}(r)+\xi_{\rm gg'}^{\rm (h2)}(r)\:,
\end{equation}
where the ``one-halo term'' $\xi_{\rm gg'}^{\rm (h1)}$ contains the contribution from galaxy pairs within the same halo and the ``two-halo term'' $\xi_{\rm gg'}^{\rm (h2)}$ from pairs within different halos. For scales which are much larger than the typical extension of a halo ($\lesssim 1\:h^{-1}$ Mpc) the one-halo term can be neglected and the correlation function becomes approximately equal to the two-halo term. In this paper, we will consider only the linear regime where the one-halo term is negligible.

The two-halo term can be approximated by
\begin{equation}\label{eq:two-halo-term}
\xi_{\rm gg'}^{\rm (h2)}(r) \simeq b_{\rm g}b_{\rm g'} \:\xi_{\rm lin}(r)\:,
\end{equation}
where we have introduced the linear bias $b_{\rm g}$ and $b_{\rm g'}$ for the two galaxy samples, respectively, being defined as
\begin{align}
b_{\rm g} &= \int \frac{d\bar n_{\rm h}}{dM}(M) b(M) \frac{\langle N_{\rm g}|M\rangle}{\bar n_{\rm g}} \: dM\\
b_{\rm g'} &= \int \frac{d\bar n_{\rm h}}{dM}(M) b(M) \frac{\langle N_{\rm g'}|M\rangle }{\bar n_{\rm g'}} \: dM\label{eq:b_g'}
\end{align}
with $d\bar n_{\rm h}/dM$ being the mass function of the DM halos and $\langle N_i |M \rangle$, $i = \{g, g'\}$, the mean number of galaxies in halos of mass $M$ determined by the corresponding HOD. In Eq.~(\ref{eq:two-halo-term}) we have neglected the extensions of the halos and formally just placed all galaxies at the centers of the corresponding halos, which is a good approximation for scales much larger than the extension of the halos.

The specific expressions for the galaxy (``$gg$'') autocorrelation and the group-galaxy  (``$Gg$'') cross-correlation function, which we will need for our analysis, are immediately obtained by means of Eqs.~(\ref{eq:two-halo-term})-(\ref{eq:b_g'}) as
\begin{align}
\xi_{\rm gg}(r) &\simeq b_{\rm g}^2 \:\xi_{\rm lin}(r)\label{eq:galaxy_auto_correlation}\\
\xi_{\rm Gg}(r) &\simeq  b_{\rm G}b_{\rm g} \:\xi_{\rm lin}(r)\:,\label{eq:galaxy_halo_correlation}
\end{align}
with
\begin{align}
b_{\rm g} &= \int \frac{d\bar n_{\rm h}}{dM}(M) b(M) \frac{\langle N_{\rm g}|M\rangle}{\bar n_{\rm g}}\: dM\\
b_{\rm G} &= \kfrac{\int f_{\rm G}(M) \frac{d\bar n_{\rm h}}{dM}(M) b(M) \: dM}{\int f_{\rm G}(M) \frac{d\bar n_{\rm h}}{dM}(M)\:dM}\:,\label{eq:linear_bias_groups}
\end{align}
where $f_{\rm G}(M)$ is the completeness of the group sample with respect to the total halo population.

\section{Data}\label{sec:selected_sample}

In this section, we describe the data that were used for the analysis in this paper. We summarize in turn the zCOSMOS survey from which the data are taken, the properties of the zCOSMOS group catalog, and the construction of realistic mock data samples. Finally, we describe the group and galaxy samples that were adopted for our correlation function analysis.

\subsection{zCOSMOS survey}

zCOSMOS \citep[][Lilly et al.~2012 in preparation]{lilly2007,lilly2009} is a deep spectroscopic galaxy survey on the 1.7 deg$^2$ of the COSMOS field \citep{scoville2007} which utilized about 600 hr of ESO Very Large Telescope service mode observations. It is divided up into two parts, ``zCOSMOS-bright'' and ``zCOSMOS-deep'', whereby this work is entirely based on the bright part, which is now complete and contains spectra in the redshift range $0.1 \lesssim z \lesssim 1.4$ for about 20,000 objects taken using the VIMOS spectrograph.

The target selection in zCOSMOS-bright is essentially magnitude limited by $15 \leq I_{\rm AB} \leq 22.5$. The slits were assigned to the targets such that for each mask the number of slit assignments on each of the four VIMOS quadrants was maximized except for some X-ray and radio objects which were observed at high priority. Since there were two masks per pointing and the pointings were overlapping with centers differing by the size of a quadrant, there were finally eight passes for the central field, four at the borders, and two at the corners. About 2\% of these spectra were taken for secondary objects, i.e., objects that were not target objects but serendipitously ended up in the slits. They are not only very helpful for estimating the accuracy and verification rate of redshifts, but also compensate for the bias against close pairs due to slit constraints \citep{deravel2011,kampczyk2011}. After removing less reliable redshifts (i.e., confidence classes 0, 1.1, 2.1 and 9.1; see \citealt{lilly2009}) and spectroscopic stars, we end up with a high-quality redshift galaxy sample containing about 16,800 objects within the area $149.47^\circ \lesssim \alpha \lesssim 150.77^\circ$ and $1.62^\circ \lesssim \delta \lesssim 2.83^\circ$. From multiply observed objects the spectral verification rate for this sample is about 99\% and the redshift accuracy about 100 km s$^{-1}$.

The spatial sampling rate (SSR), i.e., the fraction of objects of the magnitude-limited target catalog whose spectra were observed as a function of $(\alpha, \delta)$, is shown in Figure 1 of K12.  In the design of zCOSMOS there is a central region ($\alpha = 150.12\pm 0.54^\circ$ and $\delta = 2.22\pm 0.46^\circ$) with substantially higher SSR than in a region around the borders, and we will restrict our analysis to this more highly sampled region. However, even within this central region, the SSR is not completely uniform, but exhibits some stripes due to the placement of slits in the masks. It is obviously very important to take these into account especially on the smaller scales. Additionally to the SSR we have to consider the redshift success rate (RSR), which is the fraction of observed spectra yielding successfully measured redshifts (see Figs.~2 and 3 of \citealt{lilly2009}), since this affects the redshift distribution $N(z)$ of the galaxy sample.  The SSR and RSR can be assumed to be uncorrelated so that by multiplying them we obtain for each galaxy the sampling rate with respect to an ideal survey. The central area of zCOSMOS-bright has a mean completeness of 56\%. 

\begin{figure}
	\centering
	\includegraphics[width=0.46\textwidth]{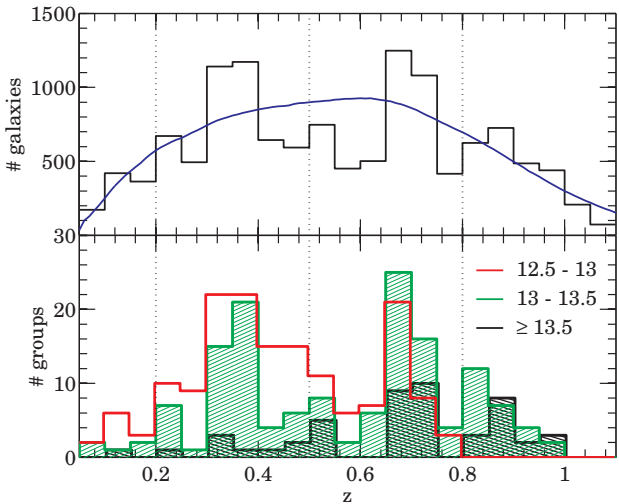}
	\caption{Number of galaxies and groups as a function of redshift. \emph{Upper panel:} distribution of zCOSMOS galaxies in the central area. The blue line shows the smoothed distribution that was used for the production of the ``randoms'' (see the text). \emph{Lower panel:} distribution of the zCOSMOS groups that were used for the cross-correlation analysis. The histograms correspond to different mass bins as indicated by the labels in units of $\log(M/M_\odot)$. The groups in the range $12 \leq \log(M/M_\odot) \leq 12.5$ are omitted for clarity. The three dotted vertical lines mark the two adopted redshift bins.\label{fig:n_of_z}}
\end{figure}

The distribution of zCOSMOS galaxies in redshift is shown in Figure \ref{fig:n_of_z}. 
There are two prominent features at redshifts $\sim\! 3.5$ and $\sim\! 0.7$. The large overdensity at high redshift is additionally contrasted by strong underdensities at $z \sim 0.6$ and $z \sim 0.8$ making the structure at $z \sim 0.7$ dominant over the redshift range $0.5 \leq z \leq 0.8$.  

\subsection{Group catalog}

The zCOSMOS 20k group catalog is described in detail in K12. The basic catalog was produced by a Friends-of-Friends (FOF) multi-run scheme \citep[see][]{knobel2009} in which we successively used different group-finding parameters, optimized for different richness groups. Based on realistic mock catalogs, the completeness and purity of groups with three or more observed members are both about 83\% with respect to all groups that should have been detectable within the survey. More than 75\% of the cataloged groups should exhibit a one-to-one correspondence to a real group in the sky (see Figs.~3 and 5 of K12).  

The purity of the group catalog can be further enhanced by selecting only the groups that are also independently detected by the Voronoi-Delaunay method (VDM). The set of FOF groups that are independently identified by VDM such that the corresponding reconstructed FOF and VDM groups exhibit a one-to-one correspondence (or two-way match, ``2WM'') is called the ``GRP$_2$ subcatalog'', which we will use for our analysis. Its purity is $\simeq\!85\%$ for groups with at least three observed members, but the completeness is lower.

Obtaining reliable observational estimates for the dynamical masses of individual groups is hard.  This is why we introduced the ``fudge masses'' $M_{\rm fudge}$, which are mass estimates based on the richness of the groups and calibrated against the mock catalogs. That is, for a given group at redshift $z$ we correct the observed spectroscopic richness for the SSR and RSR and assign it the average real halo mass of 2WM mock groups (see below) with the same corrected richness at the same redshift. The scatter between the fudge masses and the masses of the corresponding haloes in the mock catalogs is about 0.3 dex for mock groups with three members and drops to about 0.15 dex for mock groups with more than 10 members (see Tab.~7 of K12). The analysis in this paper aims to relate the fudge masses to the clustering strength of the corresponding groups.

\subsection{Mock catalogs}

We utilize the same mock catalogs that were used in K12 to produce the zCOSMOS group catalog, and for details we refer to that paper. The mock catalogs are adapted from the COSMOS mock light cones \citep{kitzbichler2007} which are based on the Millennium DM $N$-body simulation \citep{springel2005} run with the cosmological parameters $\Omega_{\rm m} = 0.25$, $\Omega_\Lambda = 0.75$, $\Omega_{\rm b} = 0.045$, $h = 0.73$, $n_{\rm s} = 1$, and $\sigma_8 = 0.9$. The semi-analytic recipes for populating the DM halos with galaxies are that of \cite{croton2006} as updated by \cite{delucia2007}. There are 24 essentially independent light cones, each covering an area of $1.4\ \rm{deg} \times 1.4\ \rm{deg}$ with an apparent magnitude limit of $r \leq 26$ and a redshift range of $z \lesssim 7$.

The mock catalogs were created to resemble as much as possible the actual 20k sample. That is, we applied the SSR and RSR of the 20k sample to the light cones by randomly removing galaxies from the magnitude limited sample according to their completeness, and we introduced a Gaussian redshift measurement error with $\sigma_z = 100(1+z)/c\ \rm{km\ s^{-1}}$ for each galaxy.

The mock group catalogs were then produced in exactly the same way as the actual group catalog was produced from the zCOSMOS data (see K12). Since within the simulations we know which galaxies belong to which groups, we make the differentiation between ``real groups'' in the mock catalogs and the ``reconstructed groups'' that are identified by the groupfinder. Reconstructed mock groups with a 2WM to real mock groups (i.e., groups that are properly identified) are called ``2WM groups''.
The statistical properties, such as purity and completeness, of the reconstructed mock groups, as determined by comparisons with the real mock groups, are then assumed to be indicative of the statistical properties of the actual identified groups in the zCOSMOS data.

\subsection{Data samples}

We restrict the analysis in this paper to the central region of the zCOSMOS field which contains about 13,000 galaxies and has a mean completeness of $56\%$ with respect to a fully magnitude-limited survey. The central region is also where most of the groups are found, where they have the highest quality, and where they are the least affected by the border of the survey. We consider in the following two redshift bins: one at high redshift $0.5 \leq z \leq 0.8$ and one at low redshift $0.2 \leq z \leq 0.5$.

It should be noted that the transverse size of the survey is much larger for the high-redshift bin than for the low-redshift bin. At $z = 0.5$ the transverse size of the survey along the declination $\Delta \delta = 0.92^\circ$ is about $22\:h^{-1}$ Mpc, while at redshift $z = 0.2$ it is only about $9\:h^{-1}$ Mpc. These lengths define the largest transverse scales for which we can hope to reliably measure the correlation function. Due to the larger transverse size at high redshift, the total volume of the survey is substantially larger resulting in a higher number of massive groups compared with the low-redshift bin.

Since the completeness of the group samples is not crucial for our analysis, we optimize our group sample for purity by adopting the GRP$_2$ subcatalog including all groups with at least three observed members. At high redshift, we divide the group population into bins of 0.5 dex in fudge mass $M_{\rm fudge}$. At low redshift, the correlation signal is much more difficult to measure due to the relatively small volume. For this reason to have enough groups in each mass bin we divide the group population into two mass bin separated by $M_{\rm fudge} = 10^{13} M_\odot$. The resulting group samples are summarized in Table \ref{tab3:group_sample} and the redshift distribution of the groups is shown in Figure \ref{fig:n_of_z}.

To cross-correlate the groups with the galaxies, we use the zCOSMOS 20k galaxy sample in the same area and in the same redshift bins. To check that the estimated bias of the groups does not depend on the adopted galaxy sample, we produce a magnitude-limited and a volume-limited galaxy sample for each redshift bin. The volume-limited samples are obtained by selecting the galaxies in absolute magnitude $M_{\rm b}$ as
\begin{equation}
M_{\rm b} \leq M_{\rm b,lim} - z\:,
\end{equation}
where the redshift $z$ is to take into account, at least approximately, the general evolution of galaxy luminosity since redshift $z \sim 1$. The absolute magnitude limit for the high-redshift bin is $M_{\rm b,lim} = -20$ and for the low-redshift bin $M_{\rm b,lim} = -19$. The galaxy samples are summarized in Table \ref{tab3:galaxy_sample}.
\begin{deluxetable}{ccccccc}
\tablewidth{0pt}
\tablecaption{Group samples}
\tablehead{
	\colhead{} & \multicolumn{3}{c}{$0.5 \leq z \leq 0.8$} & \colhead{} & \multicolumn{2}{c}{$0.2 \leq z \leq 0.5$} \\ 
\cline{2-4} \cline{6-7} \\ 	
	\colhead{Mass bins} &
  \colhead{12.5-13} &
  \colhead{13-13.5} &
  \colhead{$\geq\! 13.5$} &&
  \colhead{$\leq \! 13$} &
  \colhead{$\geq \! 13$}
  }
\startdata
No.~of groups & 53 & 64 & 24 && 140 & 60
\enddata
\tablecomments{The masses are in units of $\log(M/M_\odot)$.}

\label{tab3:group_sample}
\end{deluxetable}
\begin{deluxetable*}{cccccc}
\tablewidth{0pt}
\tablecaption{Galaxy samples}
\tablehead{
\colhead{} & \multicolumn{2}{c}{Magnitude-limited} & & \multicolumn{2}{c}{Volume-limited} \\ 
\cline{2-3} \cline{5-6} \\ 
	\colhead{}&
  \colhead{Selection} &
  \colhead{No.~of  galaxies}&
  \colhead{} &
  \colhead{Selection} &
  \colhead{No.~of  galaxies}
  }
\startdata
$0.2 \leq z \leq 0.5$ & $I_{\rm AB} \leq 22.5$ & 4712 & & $M_{\rm B}(z) = -19 - z$ & 2666 \\
$0.5 \leq z \leq 0.8$ & $I_{\rm AB} \leq 22.5$ & 4445 & & $M_{\rm B}(z) = -20 - z$ & 2212
\enddata

\label{tab3:galaxy_sample}
\end{deluxetable*} 

%\newpage
\section{Correlation function estimation}\label{sec_estimation_method}

In comoving space the correlation function is just a function of the separation $r = |\bbv x_2-\bbv x_1|$. However, in redshift space $\bf s$ the peculiar velocities of groups and galaxies distort the correlation function along the line of sight, while perpendicular to the line of sight it remains unaffected. Thus we have to deal with the ``projected correlation function''
\begin{align}
w(r_p) = \int_{-\infty}^{\infty} \xi(|\bbv s_\parallel|,|\bbv s_\perp|)\:ds_\parallel = 2 \int_{0}^{\infty} \xi(\pi,r_p)\:d\pi\:,
\end{align}
where we set $\pi = |\bbv s_\parallel|$ and $r_p = |\bbv s_\perp|$ for the decomposition $\bbv s = \bbv s_\parallel + \bbv s_\perp$ of a redshift-space vector $\bbv s$ into its components along and perpendicular to the line of sight, respectively. %$w_p(r_p)$ should be independent of the details of the redshift space distortions.
We will refer to both $\xi$ and $w$ as the ``correlation function'' if it is clear from the context which one is meant.

To estimate the autocorrelation function for a galaxy sample, the number of pairs for a given separation is compared to the corresponding number of pairs $r$ and $r'$ for an uncorrelated random catalog (the ``randoms''), which is subjected to exactly the same observational selection function as the actual galaxy sample.
We use the estimator given by \cite{landy1993}
\begin{equation}\label{eq3:landy_szalay}
\xi_{\rm gg}(\pi,r_{\rm p})= \frac{N_{\rm gg}\bar n_{\rm g}^{-2} + N_{\rm rr}\bar n_{\rm r}^{-2} - N_{\rm gr}\bar n_{\rm g}^{-1}\bar n_{\rm r}^{-1}}{N_{\rm rr} \bar n_{\rm r}^{-2}}\:,
\end{equation}
where $N_{\rm gg}$ is the number of galaxy-galaxy pairs ($gg$), $N_{\rm rr}$ is the number of random-random pairs ($rr$), and $N_{\rm gr}$ is the number of galaxy-random pairs ($gr$) each time with separations in the ranges $[\pi,\pi+d\pi]$ and $[r_{\rm p},r_{\rm p}+dr_{\rm p}]$. Note that only distinct pairs are counted.

For the cross-correlation estimation we could easily generalize this same estimator (\ref{eq3:landy_szalay}). However, it contains a term $N_{\rm rr}$ which would require us to produce also a random catalog for the groups. Since the selection function for the groups is expected to be more complicated than that for the galaxies (e.g., depending on their richnesses), we want to avoid producing a random catalog for the groups. For this reason we use a generalization of the estimator given by Peebles\nocite{peebles1973} (1973; see also \citealt{yang2005,wang2008})
\begin{equation}\label{eq:Gg_estimator}
\quad\xi_{\rm Gg}(\pi,r_{\rm p}) = \frac{N_{\rm Gg}}{N_{\rm Gr}}\:\frac{\bar n_{\rm r}}{\bar n_{\rm g}}-1\:, 
\end{equation}
where $N_{\rm Gg}$ is the number of group-galaxy pairs ($Gg$) and $N_{\rm Gr}$ is the number of group-random pairs ($Gr$) each time for separations in the ranges $[\pi,\pi+d\pi]$ and $[r_{\rm p},r_{\rm p}+dr_{\rm p}]$.

The randoms were distributed according to the SSR using the same selection function as for the production of the mock catalogs, which has a resolution of 1.5 arcmin. The scales affected by this resolution are $r_{\rm p} \lesssim 1\:h^{-1}$ Mpc even at the highest redshifts considered. The redshift distribution of the randoms was taken to be the (normalized) product of $d\bar N_{\rm g}(z)/dz$, which is the mean number of galaxies per redshift for either an ideal magnitude- or volume-limited sample, and the RSR (see blue line in Fig.~\ref{fig:n_of_z} for the magnitude-limited case). For the magnitude-limited sample we estimated $d\bar N_{\rm g}(z)/dz$ from the $I_{\rm AB} \leq 22.5$ magnitude-limited target catalog by means of a $V_{\rm max}$ like method using photometric redshifts and for the volume-limited sample we made the approximation $d\bar N_{\rm g}(z)/dz \propto dV(z)/dz$ with $V(z)$ the volume of the survey up to redshift $z$. We checked that the estimated bias neither depends sensitively on the resolution of the SSR mask nor on the details of the chosen redshift distribution.

For each pair of points (i.e., $gg$, $gr$, $rr$, $Gg$, and $Gr$) we estimated the separation by $\pi = |d_1-d_2|$ and $r_{\rm p} = (d_1+d_2)\tan(\theta/2)$, where $d_1$ and $d_2$ are the comoving distances to the points and $\theta$ their relative angle projected on the sky \citep{davis1983}. The binning for $\xi_{\rm gg}$ (and analog for $\xi_{\rm Gg}$) was taken to be logarithmic in both coordinates and the projected correlation functions $w_{\rm gg}$ were estimated by approximating the integration along $\pi$ by a sum over 13 logarithmic bins ranging from $0.1\:h^{-1}$ Mpc to $13\:h^{-1}$ Mpc. The effect of the choice of the integration range is discussed in Section \ref{sec:consistency}. We also checked that our result is insensitive to the implemented binning and integration.

The error bars for the correlation function are computed by the scatter of the correlation functions of the 24 mock catalogs. It was shown that the error bars of the $gg$ correlation function estimated by blockwise bootstrapping are of the same order as the scatter among the mock catalogs \citep{meneux2009,delatorre2010}. For the $Gg$ cross-correlation functions the group samples would be too small for blockwise bootstrapping and this procedure would anyway not take into account cosmic variance, which is not negligible even in the large COSMOS field. 

\section{The correlation functions}\label{sec:correlation_functions}

The measured correlation functions are shown in Figures \ref{fig:gg_correlation_function}-\ref{fig:Gg_correlation_function_low_maglim}. We have estimated the correlation functions for the magnitude- and volume-limited samples in both redshift bins for the zCOSMOS 20k sample and for each of the 24 mock catalogs. In the following, we first discuss the results for the mock catalogs to understand how the group selection affects the correlation function estimates and then we discuss the results for the actual data.

We show only the correlation functions for the magnitude-limited samples. The correlation functions for the volume-limited galaxy samples are qualitatively similar, but with slightly higher amplitude, since brighter galaxies reside predominantly in more massive halos, which are more biased. The estimated bias is however provided for both samples in the next section.

\newpage
\subsection{Mock correlation functions}

\begin{figure}
	\centering
	\includegraphics[width=0.46\textwidth]{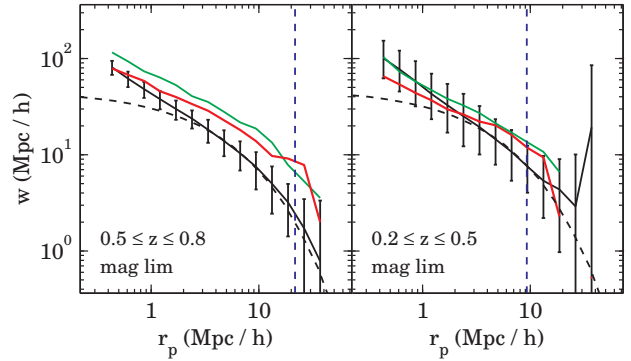}
	\caption{Autocorrelation function for the magnitude limited galaxy samples for the two redshift bins. The red line shows the 20k data and the black solid line shows the mean of the 24 mock correlation functions, where the error bars indicate their standard deviation. The dashed black line shows the linear correlation function model with adjusted normalization. For comparison the correlation function for a particular mock is also shown (green line). The vertical dashed line marks the smallest transverse box size for the corresponding redshift bin.\label{fig:gg_correlation_function}}
\end{figure}

The $gg$ correlation functions for the mock catalogs are shown in Figure \ref{fig:gg_correlation_function}.
The black solid line is the mean of the 24 mock correlation functions, where the error bars show the standard deviation among the 24 mock catalogs. The dashed black line is the linear correlation function for the cosmology of the mock catalogs, where we adjusted the normalization to that of the measured correlation function. It is obvious that for scales $\gtrsim\! 3\:h^{-1}$ Mpc we are in the linear regime and the mean of the estimated correlation functions follow very well the linear model. The vertical dashed line marks the minimal transverse box size of the survey for the corresponding redshift bin. This is the scale where we expect our estimation to fail since we cannot measure density fluctuations on scales comparable to the size of the survey. Beyond this line the error bars blow up and the different curves start to diverge from each other. This effect is even more pronounced for the $Gg$ correlation functions.

\begin{figure*}
	\centering
	\includegraphics[width=0.96\textwidth]{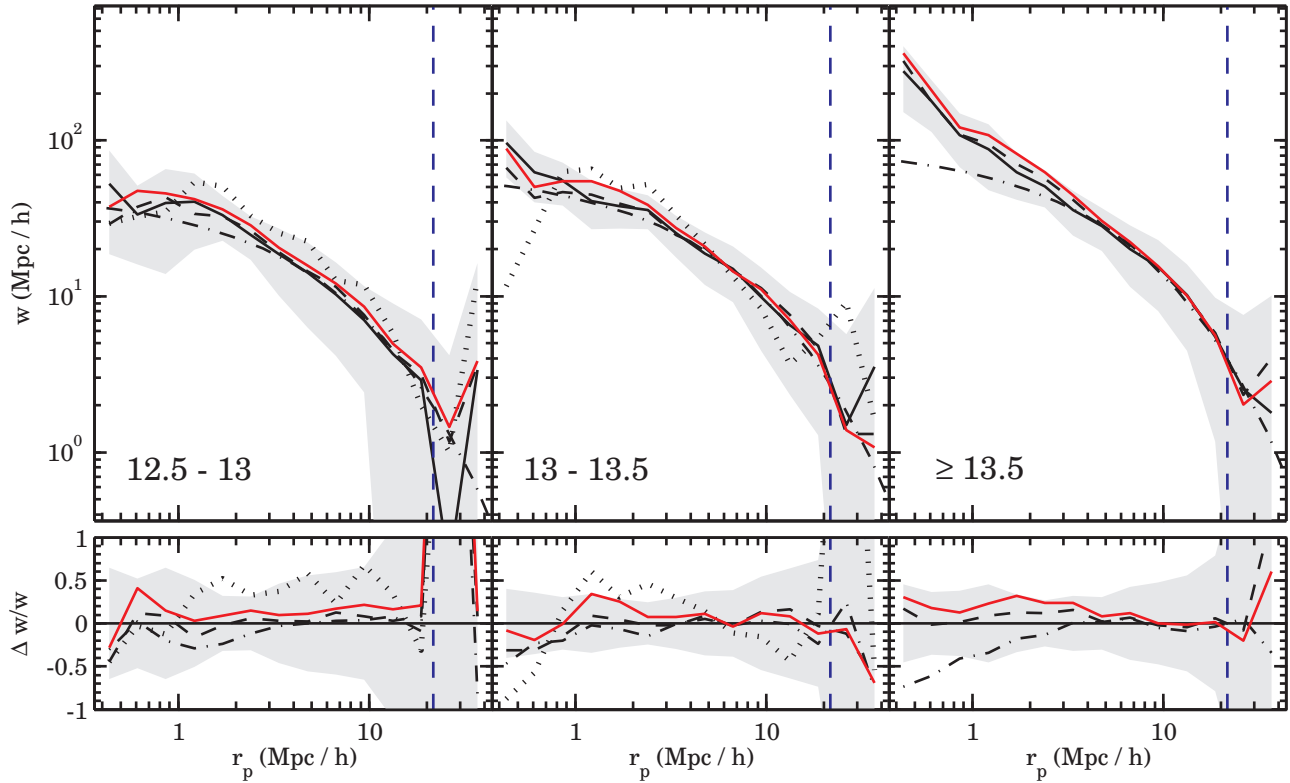}
	\caption{Group-galaxy cross-correlation functions for several mock group samples at $0.5 \leq z \leq 0.8$ using the magnitude-limited galaxy sample. The vertical panels correspond to different mass bins as indicated by the labels in units of $\log(M/M_\odot)$. In the upper panels, the curves (except dash-dotted) show the mean of the 24 mock correlation functions for different group samples, which are subsamples of the GRP$_2$ subcatalog with at least three observed members. The black solid line corresponds to real groups (i.e., real masses and real centers), the dashed line to reconstructed 2WM groups (i.e., fudge mass and estimated centers), the red solid line to all reconstructed groups, and the dotted line to the spurious groups. For comparison the dash-dotted line shows the linear correlation function model with adjusted normalization. In the lower panels, the relative difference of the correlation functions to the correlation function of the real groups (black solid line) is shown. The shaded area always corresponds to the standard deviation among the 24 samples of the real groups and the vertical dashed line marks the smallest transverse box size for this redshift bin.\label{fig:Gg_correlation_function_high_group_quality}}
\end{figure*}

\begin{figure}
	\centering
	\includegraphics[width=0.46\textwidth]{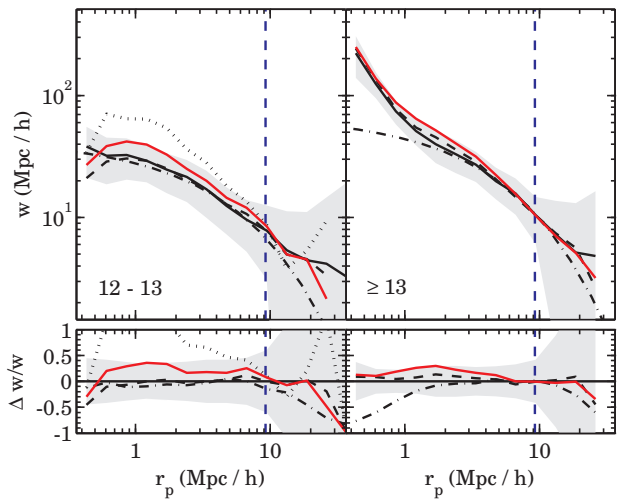}
	\caption{Group-galaxy cross-correlation function for several mock group samples at $0.2 \leq z \leq 0.5$ using the magnitude-limited galaxy sample. The different panels are for different mass bins as indicated by the labels in units of $\log(M/M_\odot)$. The lines have the same meaning as in Figure \ref{fig:Gg_correlation_function_high_group_quality}.\label{fig:Gg_correlation_function_low_group_quality}}
\end{figure}

Figures \ref{fig:Gg_correlation_function_high_group_quality} and \ref{fig:Gg_correlation_function_low_group_quality} show the mock $Gg$ correlation functions for the high- and the low-redshift bin, respectively.
We consider several group samples, which are all subsamples of the corresponding GRP$_2$ mock group catalogs with at least three observed members, and for each of these samples we show the mean of the 24 corresponding mock correlation functions.
These figures illustrate how the estimated correlation functions depend on the selection of groups. The behaviors of the different curves in both redshift bins are qualitatively similar and the main trends can be summarized as follows.

The black solid curve corresponds to the group-galaxy cross-correlation functions of the real groups in the mock (i.e., using real group positions and real halo masses within the mock catalogs) and constitutes an ideal group sample. Like for the $gg$ correlation functions they show a fair agreement to the slope of the linear correlation function (dash-dotted line) for scales $\gtrsim\!3\:h^{-1}$ Mpc. The dashed black curve is the cross-correlation function for the reconstructed 2WM mock groups (i.e., using the estimated group centers and their ``fudge'' masses). Since the estimates for the group centers feature a typical offset of $\sim \! 0.1\:h^{-1} (1+z)$ Mpc from the real group centers (see Fig.~18 of K12), they do not affect our results in the linear regime. The differences between the solid and the dashed curves are of the order $5\%$ in the linear regime, and the selection by fudge mass for 2WM groups basically has a negligible impact on the correlation function compared with the error bars.

We now look at how the non-2WM groups (i.e., the fragmented, overmerged, and spurious groups) affect the correlation function estimates. The red solid line shows the correlation function for \emph{all} reconstructed groups within the corresponding (fudge) mass range within the mock catalogs, which can be compared to the correlation functions for the actual data. They are typically about 20\% higher than the real group correlation function (given by the black solid line), but with the right slope on average. The differences to the correlation function of the 2WM groups (black dashed line) must be caused by the non-2WM groups. It can be seen that the cross-correlation function for the spurious groups (dotted lines) are typically significant higher than the other correlation functions suggesting that these spurious ``groups'' are being ``identified'' by the groupfinder in high-density environments which increases the cross-correlation of these spurious groups with the galaxies. Note that there is no dotted line shown for the highest mass bins, since there are essentially no spurious groups in our mock samples at these masses.

It is worth noting that the upturn in the cross-correlation function in the nonlinear regime is much more pronounced for the highest mass groups in the mock catalogs. This is due to the strength of the ``one-halo'' term for the richer groups. 

\subsection{zCOSMOS cross-correlation functions}

The $gg$-correlation functions for the actual data are shown in Figure \ref{fig:gg_correlation_function} (red lines). For both redshift bins they follow a nice power law, but their slopes are shallower than those for the mock catalogs. This was already pointed out by \cite{meneux2009} and \cite{delatorre2011} for the zCOSMOS 10k sample by comparing their results to correlation functions from VVDS and SDSS. In a further paper \cite{delatorre2010} suggested that the change in slope might be caused by the 10\% of galaxies residing in the most dense environments. By excluding this 10\% of high-density environment galaxies, they finally obtained consistent results.

We will follow a different route here and discuss the question whether the real 20k sample could statistically be regarded as just another mock realization with respect to the clustering of galaxies. The green lines in Figure \ref{fig:gg_correlation_function} each correspond to one of the 24 mock catalogs whose correlation function is relatively close to that of the real data. Although both green lines are by no means typical for the mock catalogs as a whole (there are about one or two like this in each redshift bin,  from the set of 24 mock catalogs) they show that the amplitude as well as the slope of the correlation functions for the actual data could be due to cosmic variance with a probability of 5\%-10\% corresponding to a $2\sigma$ effect. Compared to the mock catalogs, the COSMOS field is unusual but not exceptional.

\begin{figure}
	\centering
	\includegraphics[width=0.46\textwidth]{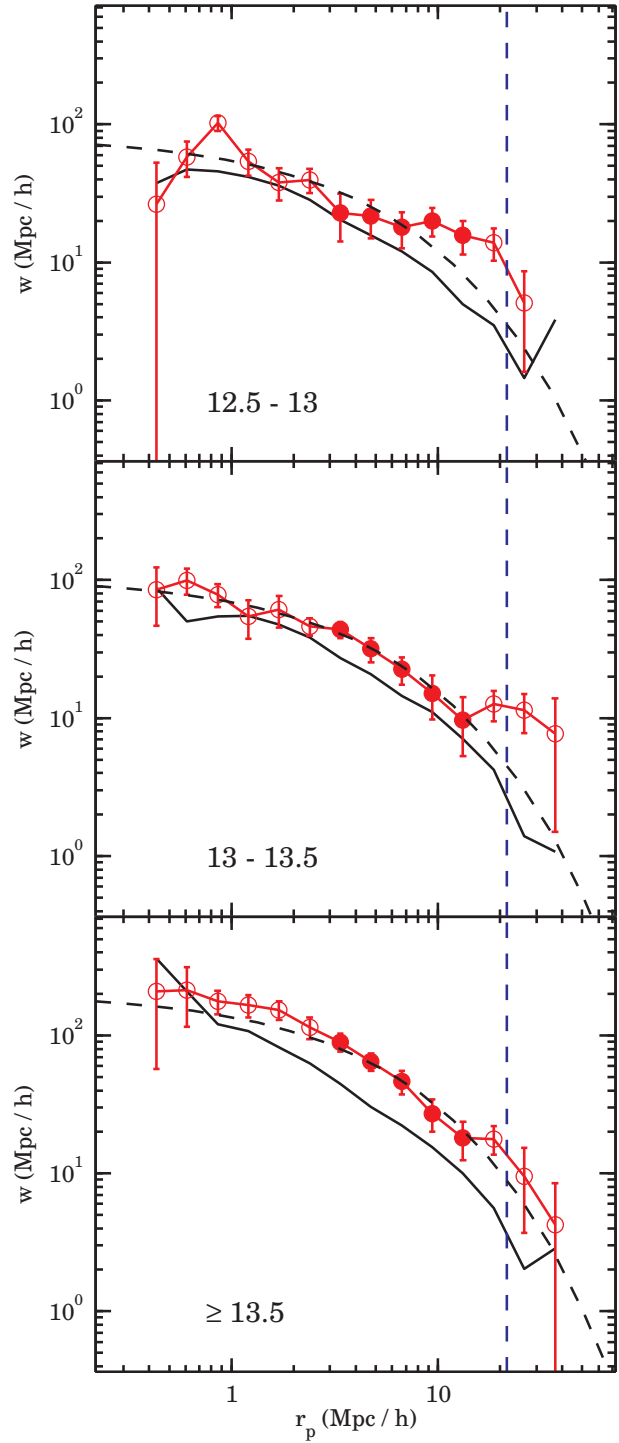}
	\caption{Group-galaxy cross-correlation functions for the zCOSMOS 20k sample (red points) at $0.5 \leq z \leq 0.8$. The different panels correspond to different mass bins as indicated by the labels in units of $\log(M/M_\odot)$. The filled points mark the linear regime and were used for the estimation of the bias. The dashed curve shows the linear model $w_{\rm lin}$ with fitted amplitude and the solid black curve shows the mean of the 24 mock cross-correlation functions (i.e., red lines in Fig.~\ref{fig:Gg_correlation_function_high_group_quality}). The error bars indicate the standard deviation of the correlation function of the 24 mock catalogs and the vertical dashed line marks the smallest transverse box size for the corresponding redshift bin.\label{fig:Gg_correlation_function_high_maglim}}
\end{figure}

\begin{figure}
	\centering
	\includegraphics[width=0.46\textwidth]{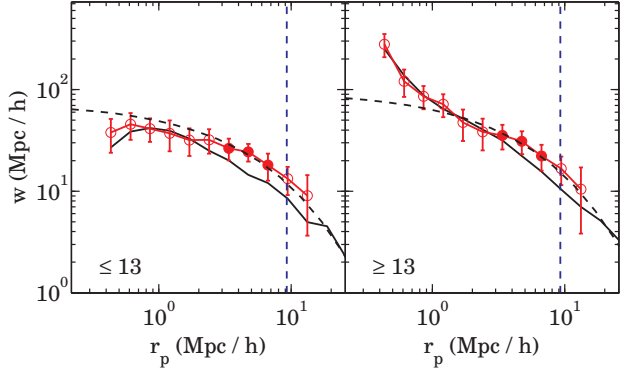}
	\caption{Group-galaxy cross-correlation functions for the zCOSMOS 20k sample (red points) at $0.2 \leq z \leq 0.5$. The different panels correspond to different mass bins as indicated by the labels in units of $\log(M/M_\odot)$. The filled points mark the linear regime and were used for the estimation of the bias. The dashed curve shows the linear model $w_{\rm lin}$ with fitted amplitude and the solid black curve shows the mean of the 24 mock cross-correlation functions (i.e., red lines in Fig.~\ref{fig:Gg_correlation_function_low_group_quality}). The error bars indicate the standard deviation of the correlation function of the 24 mock catalogs and the vertical dashed line marks the smallest transverse box size for the corresponding redshift bin.\label{fig:Gg_correlation_function_low_maglim}}
\end{figure}

The zCOSMOS $Gg$ cross-correlation functions are shown in Figures \ref{fig:Gg_correlation_function_high_maglim} and \ref{fig:Gg_correlation_function_low_maglim}.
The error bars correspond to the standard deviation among the 24 mock catalogs.
Most of the observed $Gg$ correlation functions follow nicely the linear correlation function model $w_{\rm lin}$ (dashed line, see next section) in the linear regime (filled points). For comparison also the mean $Gg$ correlation functions for all reconstructed mock groups are shown (black solid line, transposed from the red lines in Figs.~3 and 4).  While for the two lower mass bins at high redshift, the cross-correlation function is in reasonable agreement with that found in the mock catalogs, the discrepancy is quite large for the highest mass bin.  We will return to this point later.

Interestingly, the agreement between the correlation functions for the actual data (for the autocorrelation functions as well as cross-correlation functions) and for the mock catalogs is very good at small separations for all mass bins (including the highest mass bin at high redshift). This is particularly clear in Figure \ref{fig:Gg_correlation_function_low_maglim}, in which a significant detection of the one-halo term is visible for the higher mass bin. The excess in the correlation function for the actual data, relative to the mock catalogs, occurs at large scales in the linear regime.

%\newpage

\section{Mass estimation}\label{sec:bias_estimation}

In this section, we present the estimated bias for the groups. We first describe the method of estimation, then we perform a consistency test using the mock catalogs to explore possible systematics, and finally we discuss the measured masses from the bias in the actual data.

\subsection{Estimation method}

To estimate the linear group bias $b_{\rm G}$ we compare the estimated correlation functions to a model correlation function, which is a scaled version of the projected linear correlation function $w_{\rm lin}$ for the cosmology of the mock catalogs. To compute it we use the linear power spectrum
\begin{equation}
\mathcal P_{\rm lin}(k,z) = A k^{n_{\rm s}}T^2(k) D^2(z)\:,
\end{equation}
with $T(k)$ being the transfer function, $D(z)$ being the linear growth function, $n_{\rm s}$ being the spectral index, and $A$ being the normalization constant depending on $\sigma_8$. We take the fitting formula for the transfer function from \cite{eisenstein1999} using the {\tt iCosmo} software package\footnote{\url{http://www.icosmo.org}} \citep{refregier2011}. The linear correlation function $\xi_{\rm lin}$ is then obtained by the Fourier transformation
\begin{align}
\xi_{\rm lin}(r,z) =& \frac{1}{(2 \pi)^3} \int \mathcal P_{\rm lin}(k,z) \:e^{i \bbv k \bbv r}\:dk^3\\ =& \frac{1}{2 \pi^2} \int_0^\infty  k^2\:\mathcal P_{\rm lin}(k,z)\: \frac{\sin(kr)}{kr}\:dk
\end{align}
and the corresponding projected correlation function by integration along $\pi$
\begin{equation}
w_{\rm lin}(r_{\rm p},z) = 2 \int_{\pi_{\rm min}}^{\pi_{\rm max}} \xi_{\rm lin}\!\left(\sqrt{\pi^2+r_{\rm p}^2}\:,z\right)\:d\pi \:,
\end{equation}
where the integration limits are the same as for the estimated correlation functions, i.e., $\pi_{\rm min} = 0.1\:h^{-1}$ Mpc and $\pi_{\rm max} = 13\:h^{-1}$ Mpc.

With the estimated correlation functions and the theoretical model at hand, we can estimate the group bias $b_{\rm G}$ by eliminating the galaxy bias $b_{\rm g}$ from the Eqs.~(\ref{eq:galaxy_auto_correlation}) and (\ref{eq:galaxy_halo_correlation}). For each mass bin, the bias is estimated by adjusting the normalization of $w_{\rm lin}(r_{\rm p},z_{\rm eff})$ to the data points in the linear regime by means of a least-square fit, where $z_{\rm eff}$ is for either correlation function the pair-weighted effective redshift of the corresponding sample. That is, for the correlation function between the samples $G$ (or $g$) and $g$ the effective redshift is defined by
\begin{equation}
z_{\rm eff} = \frac{\sum_i  z_i N_{{\rm pairs},i} }{\sum_i N_{{\rm pairs},i}}\:,
\end{equation}
where the index $i$ runs over all objects of the sample $G$ (or $g$), $z_i$ is the redshift of the $i$th object, and $N_{{\rm pairs},i}$ is the number of pairs between the two samples that are associated to object $i$ within our maximal correlation scale. We weight the redshift (and other quantities related to a sample as a whole) by the number of pairs to account for the variation in the number of groups and galaxies with redshift, as the correlation function estimates are also pair weighted (see Eqs.~(\ref{eq3:landy_szalay}) and (\ref{eq:Gg_estimator})).
%to account for the variation in the number of groups, and galaxies, with redshift.}
The linear regime is marked by the filled points in the Figures \ref{fig:Gg_correlation_function_high_maglim} and \ref{fig:Gg_correlation_function_low_maglim}. Since the galaxy bias $b_{\rm g}$ is eliminated, the resulting group bias $b_{\rm G}$ should be independent of the choice of the galaxy sample.

Finally, we compute the mass $M_{\rm b}$ for the measured bias $b_{\rm G}$ using the fitting formula from \cite{tinker2010}, which has an accuracy of about 6\%. That is, we used the expression for $b(\nu)$ in their Eq.~(6) for the parameters in their Table 2, which is parameterized in terms of $\nu(M,z) = \delta_{\rm c}/\sigma(M,z)$, where $\delta_{\rm c} \simeq 1.686$ is the critical linear density for a spherical collapse and $\sigma(M,z)$ the linear matter variance for the mass $M$ at redshift $z$, i.e.,
\begin{equation}
\sigma(M,z) = \frac{1}{2\pi^2} \int_0^\infty k^2 \: \mathcal{P}_{\rm lin}(k,z) \: W^2(k,M)\: dk 
\end{equation}
with $W(k,M)$ being the Fourier transform for a top hat filter of scale $R = (3 M / 4 \pi \bar \rho)^{1/3}$, where $\bar \rho$ is the comoving mean matter density in the universe. This gives us the relation $b(M)$ between the linear bias and the halo mass at a given redshift, which can be numerically inverted to yield the relation $M(b)$. Thus, for a given mass $M$ we can compute the corresponding bias $b$ and vice versa.

To compare the resulting mean masses $M_{\rm b}$ that correspond to the estimates $b_{\rm G}$ to the independent fudge masses of the groups we introduce an ``effective fudge mass'' $M_{\rm eff}$ for a given group sample. We could just average the fudge masses of the groups. However, following Eq.~(\ref{eq:linear_bias_groups}), it is more correct to average the values of the bias that would be expected for the individual fudge masses, and then convert this mean bias back to a mass. Thus, we define the effective fudge mass $M_{\rm eff}$ as the mass that corresponds to the pair-weighted ``effective bias''
\begin{equation}\label{eq:effective_bias}
b_{\rm eff} = \frac{\sum_i b(M_{{\rm fudge},i}) N_{{\rm pairs},i}}{\sum_i N_{{\rm pairs},i}}\:,
\end{equation}
where the index $i$ runs over all groups within the sample, $M_{{\rm fudge},i}$ is the fudge mass for the $i$th group, and $N_{{\rm pairs},i}$ is the number of $Gg$ pairs associated with group $i$.

\newpage
\subsection{Consistency test with the mock catalogs}\label{sec:consistency}

\begin{figure}
	\centering
	\includegraphics[width=0.46\textwidth]{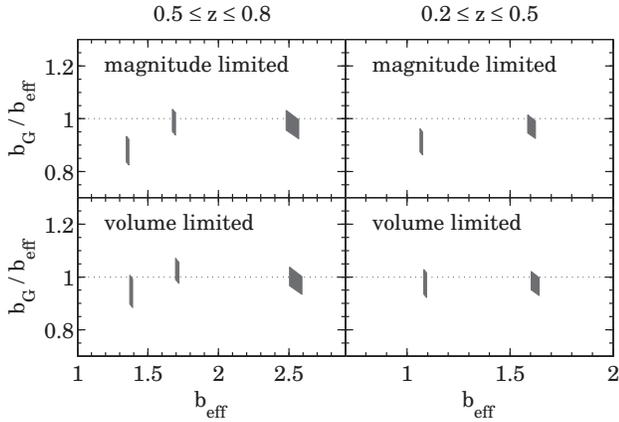}
	\caption{Consistency test for the estimation of group bias using the mock catalogs. $b_{\rm eff}$ denotes the ``true'' effective bias computed by Eq.~(\ref{eq:effective_bias}) for the real masses of 2WM mock subsamples (see the text) and $b_{\rm G}$ is the corresponding group bias obtained from the correlation function analysis. The gray areas show the $1\sigma$ regions of the means of $b_{\rm G}$ and $b_{\rm eff}$ as estimated for each of the 24 mock catalogs. The consistency between the two estimates for the magnitude- and volume-limited galaxy samples demonstrates the robustness of our method.\label{fig:bias_consistency}}
\end{figure}

To test this method, we perform an important consistency test for the subsamples of the 2WM groups within our mock group samples. Since we know the real DM halo masses for these groups (in the mock catalogs), we can estimate their ``true'' effective bias using Eq.~(\ref{eq:effective_bias}) by substituting the fudge masses by their real halo masses.
%To test this method, we perform an important consistency test for the subsample of the 2WM mock groups. \textbf{Since we know the real DM halo masses for these groups (in the mock catalogs) we can estimate their ``true'' bias using Eq.~(\ref{eq:linear_bias_groups}). For each mass bin we define the (true) effective bias as
%\begin{equation}\label{eq:effective_bias}
%b_{\rm eff} = \frac{\sum_i b(M_i) N_{{\rm pairs},i}}{\sum_i N_{{\rm pairs},i}}\:,
%\end{equation}
%where the sum runs over the index $i$ of all 2WM groups with fudge masses within the mass bin, $M_i$ is the correct DM halo mass for the $i$th group,} and $N_{{\rm pairs},i}$ is the number of $Gg$-pairs associated with group $i$.
The comparison between the group bias $b_{\rm G}$ from the correlation function analysis for these 2WM groups and their ``true'' effective bias $b_{\rm eff}$ is shown in Figure \ref{fig:bias_consistency} for all samples.
The gray parallelograms represent the estimated $1\sigma$ region for the mean of the 24 mock catalogs in each mass bin.  They take into account the scatter of $b_{\rm G}$ from the 24 mock correlation functions and also the scatter of $b_{\rm eff}$ and are these standard deviations divided by root-24.  The fact that the recovered mean bias $b_{\rm G}$ in the 24 mock catalogs is always close to the actual bias $b_{\rm eff}$ (averaged over the range of redshift and mass etc.) suggests that systematic errors (e.g., due to the uncertainty in the $b(M)$ relation) are well below the random uncertainties associated with a single mock catalog, or with a single COSMOS-like survey field. 

\begin{figure}
	\centering
	\includegraphics[width=0.46\textwidth]{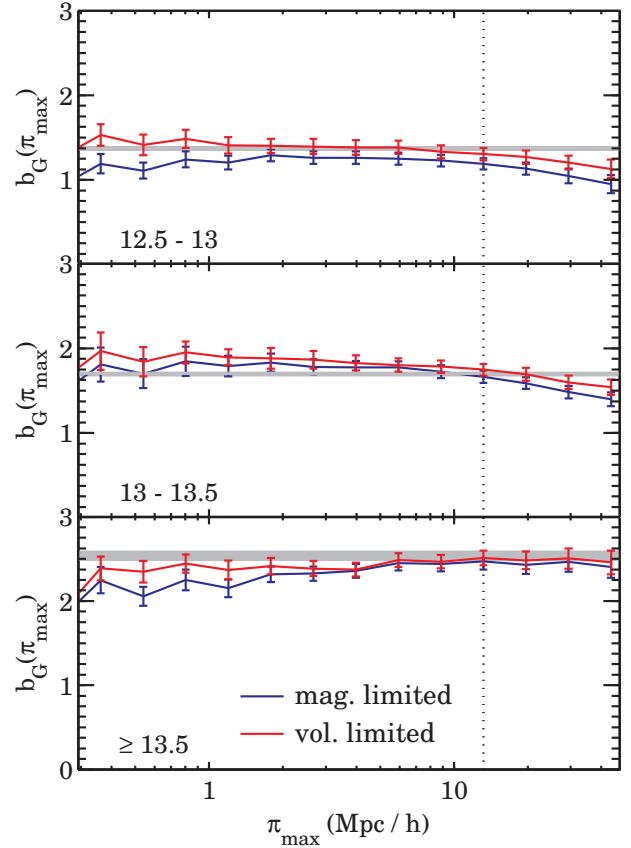}
	\caption[]{Group bias $b_{\rm G}$ as a function of the maximum integration limit $\pi_{\rm max}$ for the 2WM mock subsamples in the high redshift bin. The different panels correspond to different mass bins according to the labels in units of $\log(M/M_\odot)$. The red curve corresponds to the mean bias of the 24 mock catalogs of the volume-limited galaxy sample and the blue curve corresponds to the mean of the magnitude-limited sample. The error bars show the standard deviation of the mean and the gray horizontal bar the $1\sigma$ region of the mean of the ``true'' effective bias $b_{\rm eff}$ (see the text). The vertical dotted line marks the adopted integration limit for this paper.\label{fig:integration_limits}}
\end{figure}

We also explored (again using the 2WM mock subsamples) how the estimated bias depends on the maximum integration limit $\pi_{\rm max}$ for the different galaxy and group samples. For the high-redshift bin, the corresponding curve is shown in Figure \ref{fig:integration_limits}.
For the low-redshift bin (not shown) the trends are very similar. The curves correspond to the mean of the 24 mock catalogs and the error bars show the standard deviation of the mean. It turns out that the bias generally depends only weakly on the integration limit. The bias of the volume-limited sample (red) is systematically slightly higher than that of the magnitude-limited samples (blue) and approaches the expected bias closer (this is also obvious in Fig.~\ref{fig:bias_consistency}). Another feature is the slight decline of the bias for $\pi_{\rm max} \gtrsim 15\:h^{-1}$ Mpc, particularly for lower mass bins. For this reason we chose a maximum integration limit which is lower than the $20\:h^{-1}$ Mpc adopted in \cite{meneux2009} and \cite{delatorre2011}. The difference in the estimated bias changes, however, only about 10\% between these two integration limits, which is small compared with the error bar of the bias for a single mock catalog which is $\sqrt {24}$ times larger than the error bars shown in the Figures \ref{fig:bias_consistency} and \ref{fig:integration_limits}.

\subsection{Resulting masses and discussion}\label{sec:masses}

\begin{figure*}
	\centering
	\includegraphics[width=0.70\textwidth]{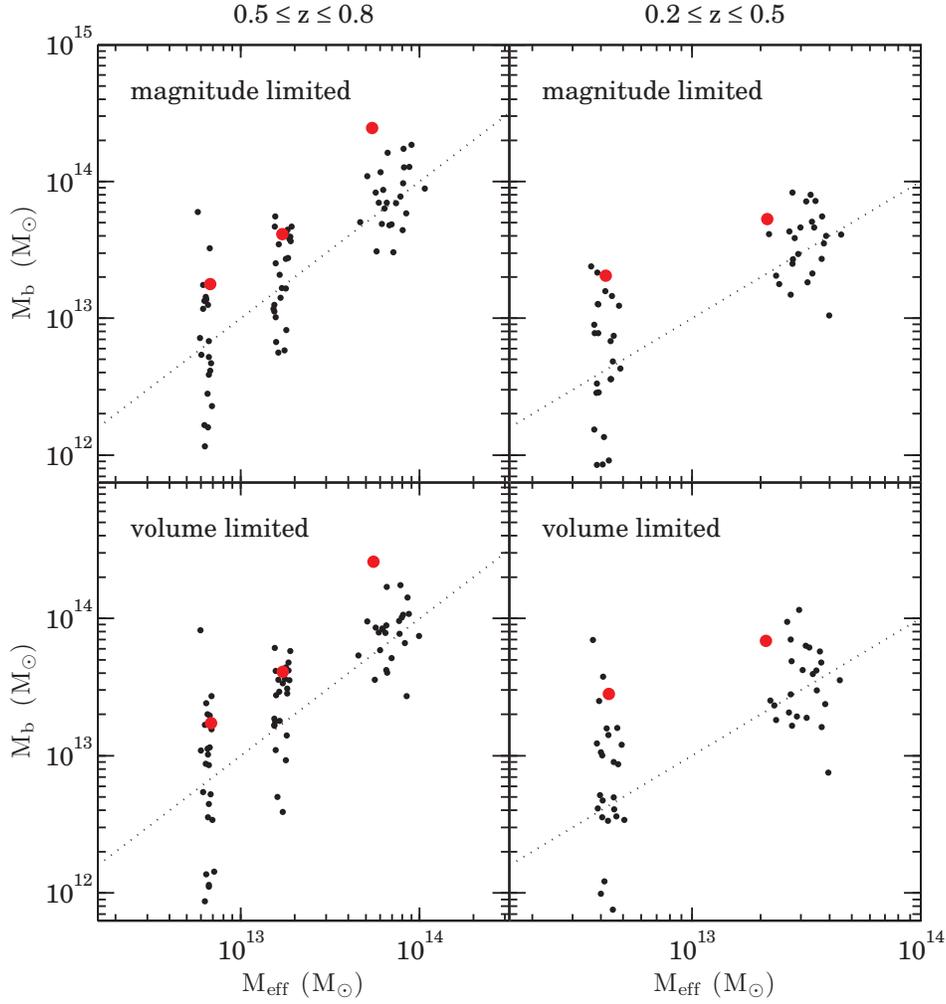}
	\caption{Estimated mean masses $M_{\rm b}$ from the correlation function analysis as a function of the effective fudge mass $M_{\rm eff}$ within the corresponding mass bins (see Tab.~\ref{tab3:group_sample}). The red points show the 20k groups and the black points correspond to each of the 24 mock catalogs. The different mass bins are well separated from each other in terms of $M_{\rm eff}$ and the galaxy sample that was used is indicated in the panels. Equality of $M_{\rm b}$ and $M_{\rm eff}$ is marked by the dashed line.\label{fig:estimated_mass}}
\end{figure*}

The resulting mean masses $M_{\rm b}$ that correspond to the estimates $b_{\rm G}$ for our group samples are shown in Figure \ref{fig:estimated_mass} as a function of the effective fudge mass $M_{\rm eff}$ defined by means of Eq.~(\ref{eq:effective_bias}).
%\textbf{Our consistency test is finally performed by comparing the resulting mean masses $M_{\rm b}$ that correspond to the estimates $b_{\rm G}$ for our group samples to the effective fudge mass $M_{\rm eff,fudge}$ for each mass bin. The effective fudge mass is the mass that corresponds to the effective (fudge) bias
%\begin{equation}
%b_{\rm eff,fudge} = \frac{\sum_i b(M_{{\rm fudge,}i}) N_{{\rm pairs},i}}{\sum_i N_{{\rm pairs},i}}\:,
%\end{equation}
%where here, in contrast to Eq. (XXX), $i$ runs over all galaxies with fudge masses in the corresponding mass bin and $M_{{\rm fudge,}i}$ is the fudge mass of the $i$th galaxy.}
The masses $M_{\rm b}$ along with $b_{\rm G}$ for the zCOSMOS groups are provided in Table \ref{tab3:bias}, where the masses and the bias assume $\sigma_8 = 0.9$, but the latter can be easily scaled to any value of $\sigma_8$ as $b_{\rm G}(\sigma_8) = (0.9/\sigma_8) b_{\rm G}(0.9)$. On the other hand, the resulting masses are relatively insensitive to $\sigma_8$ as the change in $b_{\rm G}(\sigma_8)$ is counteracted by the mass-bias relation for different $\sigma_8$. We checked that for $\sigma_8 = 0.8$, as indicated by recent measurements \citep[e.g.,][]{komatsu2011}, the resulting $M_{\rm b}$ would decrease by $\sim\!0.05$-$0.1$ dex with a stronger decline for the higher masses.%, which does not change the main trends shown in Figure \ref{fig:estimated_mass}.

The red points in Figure \ref{fig:estimated_mass} correspond to the zCOSMOS 20k groups, while the black points show, for each bin, the resulting masses for the 24 mock catalogs.
The relatively low effective fudge mass for the actual data in the high-mass bin at low redshift is due to the lack of high richness groups relative to the COSMOS light cones (see K12 for a discussion). The overall similarity between the estimated masses for the magnitude- and volume-limited galaxy samples underscores the robustness of the method and the expected independence of the analysis on the details of the galaxy sample (the magnitude- and volume-limited samples differ by a factor of about two in number, see Tab.~\ref{tab3:galaxy_sample}). The difference between the estimated masses are only about 0.02 dex for the high-redshift range and $\lesssim \! 0.15$ dex for the low redshift range (cf.~Tab.~\ref{tab3:bias}).
\begin{deluxetable*}{crrrcrr}
\tablewidth{0pt}
\tablecaption{Measured values of the bias $b_{\rm G}$ and corresponding mean masses $M_{\rm b}$ for the zCOSMOS 20k groups}
\tablehead{
	\colhead{} & \multicolumn{3}{c}{$0.5 \leq z \leq 0.8$} & \colhead{} & \multicolumn{2}{c}{$0.2 \leq z \leq 0.5$} \\ 
\cline{2-4} \cline{6-7} \\ 	
	\colhead{Mass bins} &
  \colhead{12.5-13} &
  \colhead{13-13.5} &
  \colhead{$\geq\! 13.5$} &&
  \colhead{$\leq \! 13$} &
  \colhead{$\geq \! 13$}
  }
\startdata
\sidehead{Magnitude-limited galaxy samples}

$b_{\rm G}$  & $1.64 \pm 0.36 $ & $2.12 \pm 0.33$ & $4.16 \pm 0.48$ && $1.52 \pm 0.25$ & $1.97 \pm 0.28$ \\
$M_{\rm b}$& $13.25 \pm 0.91$ & $13.61\pm 0.31$ & $14.39\pm 0.21$ && $13.45\pm 0.47$ & $13.84\pm 0.28$   \\

\sidehead{Volume-limited galaxy samples}

$b_{\rm G}$ & $1.64\pm 0.39$ & $2.14\pm 0.33$ & $4.30\pm 0.43$ && $1.39\pm 0.18$  & $1.81\pm 0.25$ \\
$M_{\rm b}$ & $13.24 \pm 0.67$ & $13.61 \pm 0.31$ & $14.41 \pm 0.20$ && $13.31 \pm 0.44$ & $13.72\pm 0.24$ 

\enddata

\tablecomments{The masses are in units of $\log(M/M_\odot)$ and we explicitly assumed $h = 0.73$ and $\sigma_8 = 0.9$.  For another choice of $\sigma_8$ the bias can be easily rescaled as $b_{\rm G}(\sigma_8) = (0.9/\sigma_8) b_{\rm G}(0.9)$. The error bars correspond to the standard deviation among the 24 mock catalogs. For the definition of the galaxy samples see Table \ref{tab3:galaxy_sample}.}

\label{tab3:bias}
\end{deluxetable*}

The linear bias $b(M)$ is expected to be a strongly nonlinear function of $\log(M/M_\odot)$. At redshift $z = 0.7$, for instance, it sharply increases for $M \gtrsim 10^{13.5} M_\odot$ and flattens for $M \lesssim 10^{12} M_\odot$ \citep[e.g.,][]{pillepich2010,tinker2010}. At lower redshift, these features move to slightly higher masses (i.e., the shift is about 0.5 dex until $z = 0$).  These features in the $b(M)$ curve will, for instance, produce non-Gaussian distribution of the indicated mass for a Gaussian observed uncertainty in $b$.  For this reason, we plot the distributions of the indicated $M_{\rm b}$  for the 24 mock catalogs instead of error bars on the mean, which would be difficult to interpret. Not surprisingly (cf.~Fig.~\ref{fig:bias_consistency}), the measured bias $b_{\rm G}$ for the mock catalogs are, nonetheless, quite symmetrically distributed around the corresponding effective bias $b_{\rm eff}$ derived from the fudge masses (see Eq.~(\ref{eq:effective_bias})), but are typically 5\%-10\% higher than $b_{\rm eff}$. This is consistent with the systematic overestimation of the correlation function for reconstructed mock groups compared to that for 2WM groups as shown in the Figures \ref{fig:Gg_correlation_function_high_group_quality} and \ref{fig:Gg_correlation_function_low_group_quality}. 

The estimated mean masses $M_{\rm b}$ for the zCOSMOS groups are broadly within the scatter of the 24 mock catalogs. They increase with fudge mass as expected by the theory of cosmic structure formation. This effect was first clearly measured by \cite{bahcall1983} using Abell clusters and recently in SDSS in the low-redshift universe with a detailed comparison to the $\Lambda$CDM cosmology \citep[e.g.,][]{berlind2006,wang2008}. The masses $M_{\rm b}$ for the zCOSMOS groups are, however, generally larger than the average $M_{\rm b}$ for the mock catalogs reflecting the high amplitudes of the correlation functions in Figures \ref{fig:Gg_correlation_function_high_maglim} and \ref{fig:Gg_correlation_function_low_maglim} (i.e., the measured bias is typically about 1$\sigma$-1.5$\sigma$ higher than the corresponding mean bias in the 24 mock catalogs). An exceptionally large bias is measured for the highest mass bin at high redshift, this being 15\% higher than the largest bias seen in the corresponding 24 mock samples. This is an about $3\sigma$ effect. Adopting a lower $\sigma_8$ of 0.8 would mitigate the discrepancy by decreasing the corresponding $M_{\rm b}$ for the actual data by  $\sim \! 0.1$ dex, but it would still be higher than in any of the 24 mock catalogs. The reason for this unusually large bias measurement might be the huge structure at redshift $z \sim 0.7$ (see Fig.~\ref{fig:n_of_z}) in the COSMOS field (see also the discussion in \citealt{meneux2009}). In fact, of the 24 groups in this mass bin, 19 lie in the redshift range $0.65 \leq z \leq 0.75$ and the remaining five groups are at $z < 0.55$. Thus, the corresponding correlation function is almost entirely dominated by the structure at $z \sim 0.7$.

Thus, we conclude that in total our finding is essentially consistent with simulations. Although this analysis provides an important overall consistency check, the rather large error bars emphasize that these correlation functions do not provide precise estimates of the halo masses of particular tracers unless very large samples in very large fields are available. There are, however, indications that the structures observed in the COSMOS field correspond to rather rare realizations of the mock catalogs. The relatively high values of the bias reflecting the large amplitudes of the measured correlation functions are in general agreement with previous studies \citep{mccracken2007,meneux2009,kovac2010a,delatorre2010,delatorre2011}, which reported unusually strong clustering in the COSMOS field.

\section{Conclusion}\label{sec:conclusion_correlation_function}

We have performed a group-galaxy cross-correlation analysis in two redshift bins (i.e., $0.5 \leq z \leq 0.8$ and $0.2 \leq z \leq 0.5$) using the high-quality zCOSMOS group sample cross-correlated with two spectroscopic zCOSMOS galaxy samples. The aim was to perform a consistency test between the clustering strength of groups and their masses that had previously been estimated on the basis of the observed richness of the groups. To compute the group bias $b_{\rm G}$ we measured the cross-correlation function between groups and galaxies and eliminated the galaxy bias $b_{\rm g}$ by also measuring the galaxy autocorrelation function. The analysis was carried out using both magnitude- and volume-limited galaxy samples to demonstrate the robustness of our estimates.

The mock catalogs are a valuable tool to explore the systematics of the methods and samples. A comparison of the cross-correlation functions for real mock groups and for all reconstructed mock groups shows that our group catalog is suited for a correlation function analysis, but that we overestimate the correlation function for real groups by about 20\% due to imperfections in the group catalog and in particular the spuriously identified groups. We also performed a consistency test with the mock catalogs to test our method and found largely consistent results between the clustering strength and the group masses.

The resulting galaxy autocorrelation and group-galaxy cross-correlation functions for the actual 20k zCOSMOS group samples follow, as expected, approximate power laws in the linear regime which assures that the estimated bias is a meaningful quantity. At high redshift the amplitude of the galaxy autocorrelation function is rather high compared with the results from the mock catalogs and at both high and at low redshift its slope is unusually shallow (for the high-redshift bin this was already noted in \citealt{meneux2009}; \citealt{delatorre2010}). We checked, however, that there are individual mock catalogs within our ensemble set of 24 which exhibit similar autocorrelation function amplitudes and slopes. This reassures that the zCOSMOS sample with respect to clustering can be regarded as a slightly untypical mock cone, where the peculiarities are simply due to cosmic variance.

The measured bias for the zCOSMOS groups increases with group richness as expected by the theory of cosmic structure formation and yields masses that are reasonably consistent with the masses derived from the richness alone, considering the scatter that is obtained from the 24 mock catalogs. Only the bias for the highest mass bin at high redshift is significantly higher than seen in any of the 24 mock catalogs, which corresponds to about a 3$\sigma$ effect. This can likely be attributed to the extremely large structure that is present in the COSMOS field at $z \sim 0.7$. The small differences between the estimated masses for the magnitude- and volume-limited galaxy samples demonstrate the robustness of our result at low and at high redshift.

In total we find overall fairly consistent results between the zCOSMOS sample and numerical simulations, although there are indications that the structures observed in the COSMOS field correspond to rare realizations of the COSMOS light cones. Our measured values of the bias are systematically larger than on average within the simulations, which reflects the unusual strong clustering in the COSMOS field that was reported in previous studies \citep{mccracken2007,meneux2009,kovac2010a,delatorre2010,delatorre2011}.

This research was supported by the Swiss National Science Foundation, and it is based on observations undertaken at the European Southern Observatory (ESO) Very Large Telescope (VLT) under the Large Program 175.A-0839.

\bibliographystyle{apj3}
\bibliography{apj-jour,bibliography}

\end{document}